\documentclass[epj]{svjour}
\usepackage{epsfig}
\usepackage{graphics}
\newcommand{\ga}{\alpha}
\newcommand{\gb}{\beta}
\newcommand{\gc}{\gamma}
\newcommand{\gd}{\delta}

\newcommand{\gl}{\lambda}
\newcommand{\gs}{\sigma}
\newcommand{\gS}{\Sigma}
\begin{document}
\title{Few-body correlations in the QCD phase diagram}
\author{Michael Beyer
}                     
%
%
\institute{Fachbereich Physik, Universit\"at Rostock, 18051 Rostock, Germany}
\date{Received: date / Revised version: date}
%
\abstract{From the viewpoint of statistical physics, nuclear matter
is a strongly correlated many-particle system. Several regimes of the 
QCD phase diagram should exhibit strong correlations. Here I focus on
three- and four-body correlations that might be important in the
phase diagram.
\PACS{
      {21.65.+f}{Nuclear matter} \and
      {21.45.+v}{Few-body systems}
     } 
} 
\maketitle
\section{Introduction}
\label{intro}
Lattice and effective model calculations provide a rich and exciting
sketch of the phase diagram of Quantum Chromodynamics (QCD).  Two
regions may coarsely be distinguished: a hadronic phase and a plasma
phase.  Since quarks are considered part of the fundamental building
blocks of matter the existence of a hadronic phase is already an
indication of strong correlations between quark-antiquark and
three-quarks. Further on, in both phases correlations lead to more interesting
phenomena, such as clustering of nucleons to form nuclei, or
superfluidity (in nuclear matter) and color superconductivity (in
quark matter). In particular in the later cases a weak residual
interaction is enough to destabilize the ground state (just as is the
case for the formation of Cooper pairing).  These investigations are
usually based on the study of two-particle correlations. There are
reasons to go beyond two-particle correlations, e.g.:
\begin{itemize}
\itemsep=0pt
\item Particle production even in a dense environment such as deuteron
  formation in a heavy ion reaction, need a third particle to conserve
  energy-momentum~\cite{Kuhrts:2001zs}.
\item To study the properties of
  $\alpha$-particles~\cite{Beyer:2000ds} or determine the critical
  temperature of a possible $\alpha$-particle
  condensate~\cite{alpha,Tohsaki:2001an,Suzuki:2002me}
  needs an in-medium four-body equation.
\item Recent results in the Hubbard model indicate, that
  three-particle contributions may lead to a different (lo\-wer)
  critical temperature compared to the simple Thouless
  criterion~\cite{letz}. Question of this type have not been addressed
  for nuclear matter.
\item The chiral phase transition is often discussed along with a
  confinement-deconfinement transition based on investigating mesons
  (quark-antiquark states), see e.g. Ref.~\cite{Blaschke}. Does this
  transition happen for nucleons (three-quark states) at the same 
density/temperature?
\end{itemize}
To investigate these issues a first step is to develop and solve
proper effective in-medium three- and four-body equations that are
valid at {\em finite temperatures and densities} analogous to the
Feynman-Galitskii or Bethe-Goldstone equations~\cite{fet71}.  This has
been achieved in the past for the nonrelativistic
problem~\cite{Beyer:2000ds,Beyer:1996rx,Beyer:1999tm,Beyer:1999zx}.
These equations have been derived on the basis of statistical Green
functions~\cite{fet71}. The Green functions have been decoupled
utilizing a cluster expansion, see e.g.~\cite{duk98}.  To tackle these
questions in the (deconfined) quark-phase, in addition, such in-medium
few-body equations must obey special relativity. As chiral symmetry
breaking is presumably restored (up to small current masses) the
quarks may become very light objects and therefore relativistic
effects should play a larger role than for isolated systems.  Here,
relativity is realized using the light front form of relativistic
dynamics~\cite{Dirac:49}.  First results are given in
Refs.~\cite{Beyer:2001bc,Mattiello:2001vq} on the
confinement-deconfinement (Mott) transition. 

\section{Theory}
We use Dyson equations to tackle the many-particle problem, see e.g.
Ref.~\cite{duk98}. This enables us to decouple the hierarchy of Green
functions. The Dyson equation approach used here is based on two
ingredients: i) all particles of a cluster are taken at the same
global time ii) the ensemble averaging for a cluster is done for an
uncorrelated medium.  The resulting decoupled Green functions may be
economically written as resolvents in the $n$-body space, where
$n=2,3,4,\dots$ is the number of particles in the considered cluster.

The solution of the one-particle problem in Hartree-Fock approximation leads to the following quasi-particle energy
\begin{equation}
\varepsilon_1 = \frac{k^2_1}{2m_1}+\sum_{2}V_2(12,\widetilde{12})f_2
\simeq \frac{k^2_1}{2m_1^{\rm eff}}+\gS^{\rm HF}(0).
\end{equation}
The last equation introduces the effective mass that is a valid
concept for the rather low densities considered here and $\mu^{\rm
  eff}\equiv \mu - \gS^{\rm HF}(0)$.  The Fermi function $f_i\equiv
f(\varepsilon_i)$ for the $i$-th particle is given by
\begin{equation}
f(\varepsilon_i) = \frac{1}{e^{\gb(\varepsilon_i - \mu)}+1}.
\end{equation}
The resolvent $G_0$ for $n$ noninteracting quasiparticles is
\begin{equation}
G_0(z) = (z-  H_0)^{-1}
N \equiv R_0(z) N,\qquad H_0 = \sum_{i=1}^n \varepsilon_i
\end{equation}
where $G_0$, $H_0$, and $N$ are formally matrices in $n$ particle
space.  The Matsubara frequency $z_\gl$ has been analytically
continued into the complex plane, $ z_\gl\rightarrow
z$~\cite{fet71}. The Pauli-blocking for $n$-particles is
\begin{equation}
N=\bar f_1\bar f_2 \dots \bar f_n
\pm f_1f_2\dots f_n,\qquad\bar f=1-f
\end{equation}
where the upper sign is for Fermi-type and the lower for Bose type
clusters. The full
resolvent $G(z)$ is given by
\begin{equation}
G(z)=(z-H_0-V)^{-1}{N}
, \qquad
V\equiv \sum_{\mathrm{pairs}\;\ga} N_2^{\ga}V_2^{\ga}.
\end{equation}
Note that $V^\dagger\neq V$.
For the two-body case as well as for a two-body subsystem embedded in
the $n$-body cluster the standard definition of the $t$ matrix leads
to the Feynman-Galitskii equation for finite temperature and
densities~\cite{fet71},
\begin{equation}
T_2^\ga(z) =   V_2^\ga + 
 V_2^\ga  N^\ga_2 R_0(z)  T_2^\ga(z).
\label{eqn:T2}
\end{equation}
Introducing the  Alt Grassberger Sandhas (AGS)~\cite{alt67}
transition operator $U_{\alpha\beta}(z)$
the effective inhomogeneous in-medium AGS equation reads
\begin{equation}
U_{\alpha\beta}(z)= (1-{\delta}_{\alpha\beta})R^{-1}_0(z)+
\sum_{\gamma\neq \alpha}
{N^\gc_2}
T_2^\gamma(z)R_0(z)
U_{\gamma\beta}(z).
\label{eqn:T3}
\end{equation}
The homogeneous in-medium AGS equation uses the form factors defined by
\begin{equation}
|F_\gb\rangle\equiv\sum_\gc\bar\gd_{\gb\gc} { N_2^\gc}  V_2^\gc 
|\psi_{B_3}\rangle
\end{equation}
to calculate the bound state $\psi_{B_3}$
\begin{eqnarray}
|F_\ga\rangle
&=&\sum_\gb \bar\gd_{\ga\gb}  
{N_2^\gb} T_2^\gb(B_3) R_{0}(B_3)|F_\gb\rangle.
\end{eqnarray}
Finally, the four-body bound state is described by
\begin{equation}
|{\cal F}^\gs_\gb\rangle=\sum_{\tau\gc} \bar\gd_{\gs\tau}
U^\tau_{\gb\gc}(B_4)  R_0(B_4) { N_2^\gc} 
T_2^\gc(B_4) R_0(B_4) |{\cal F}^\tau_\gc\rangle,
\end{equation}
where $\ga\subset\gs,\gc\subset\tau$ and $\gs,\tau$ denote the
four-body partitions. The two-body input is given in (\ref{eqn:T2})
and the three-body input by (\ref{eqn:T3}). Note that, although we
have managed to rewrite the above equations in a way close to the ones
for the isolated case, they contain all the relevant in-medium
corrections in a systematic way, i.e. correct Pauli-blocking and self
energy corrections. The numerical solution requires some mild
approximations that are however well understood in the context of the
isolated few-body problem.

\section{Results}
An experiment to explore the equation of state of nuclear matter is
heavy ion collisions at various energies. Here we focus on
intermediate to low scattering energies and compare results to a
recent experiment $^{129}$Xe+$^{119}$Sn at 50 MeV/A by the INDRA
collaboration~\cite{INDRA}. A microscopic approach to tackle the heavy
ion collision is given by the Boltzmann equation for different particle
distributions and solved via a Boltzmann Uehling
Uhlenbeck (BUU) simulation ~\cite{dan91,Dan92}. The reaction rates 
appearing in the collision integrals are {\em a priori} medium dependent.
However, previously this medium dependence has been neglected.  Within
linear response theory for infinite nuclear matter the use of
in-medium rates leads to faster time scales for the deuteron life time
and the chemical relaxation time as has been shown in detail in
Refs.~\cite{Beyer:1997sf,Kuhrts:2000jz}. This faster time scales
should have consequences for the freeze out of fragments.

\begin{figure}[b]
\begin{minipage}{0.49\textwidth}
\epsfig{figure=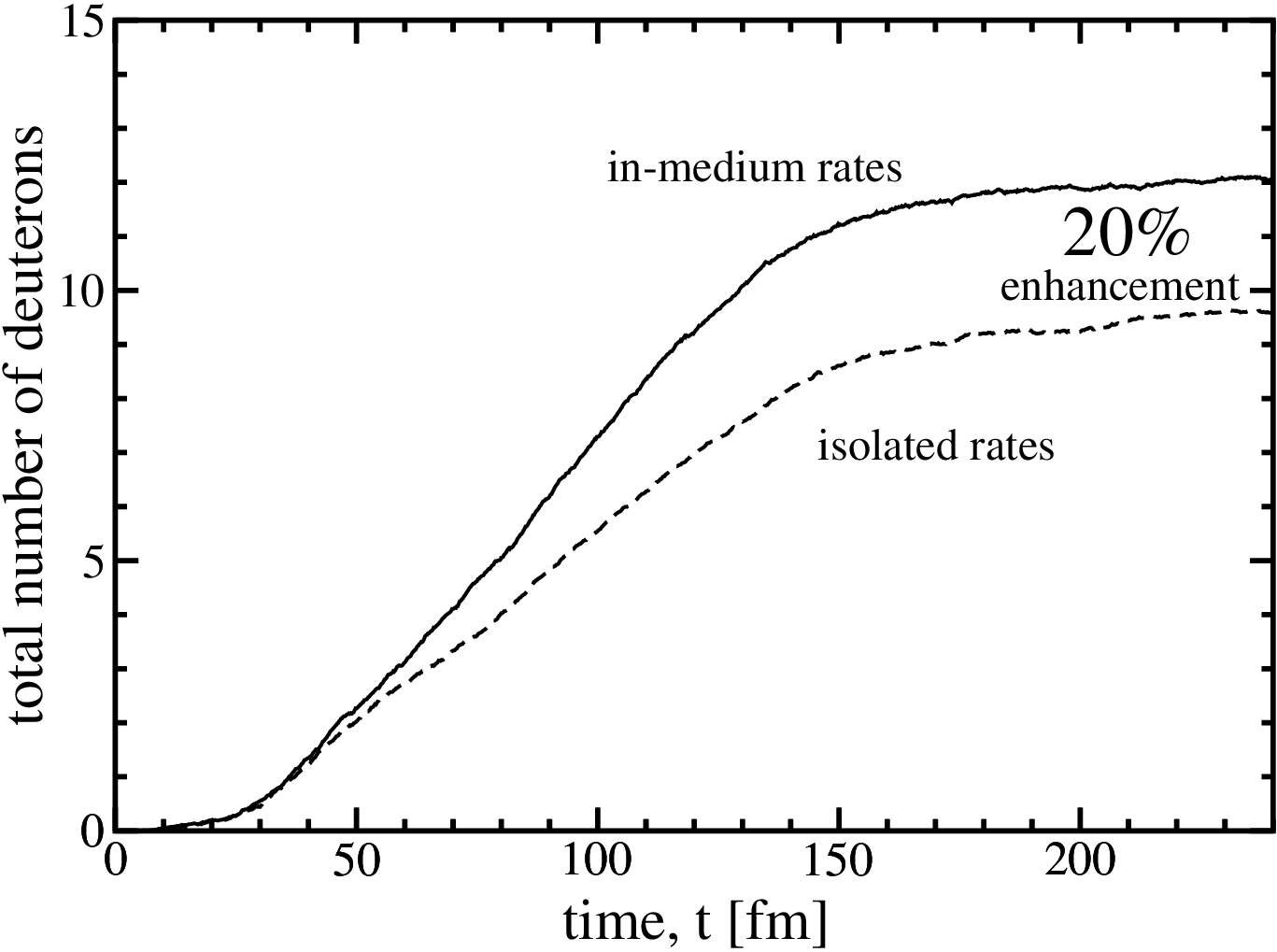,width=0.9\textwidth}
\caption{\label{fig:PLT} 
  BUU simulation of the deuteron formation during the central collision of
  $^{129}$Xe+$^{119}$Sn at 50 MeV/A.}
\end{minipage}\hfill
\begin{minipage}{0.49\textwidth}
  \epsfig{figure=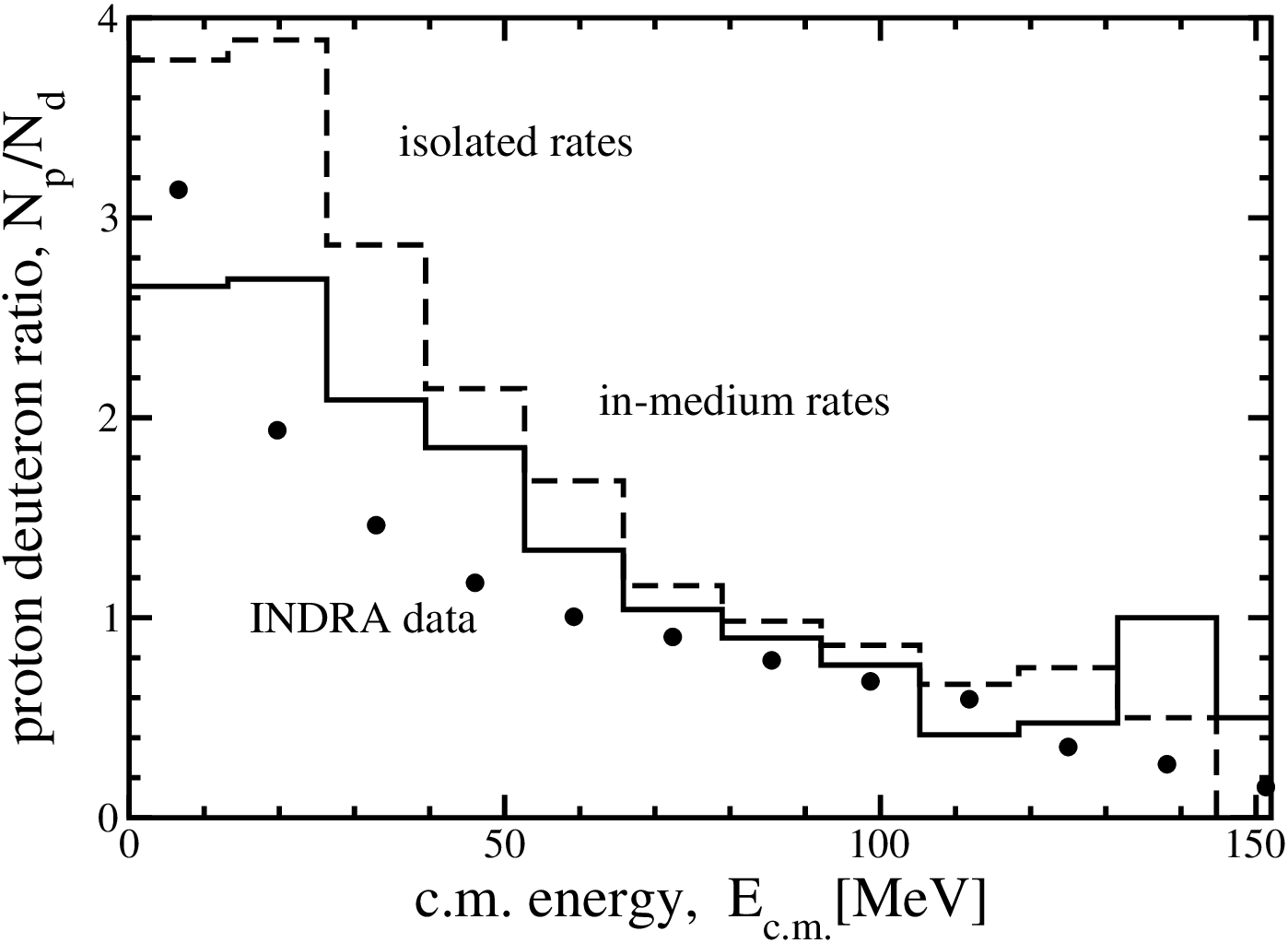,width=0.9\textwidth}
\caption{\label{fig:pd} 
  Ratio of proton to deuteron numbers as a function of c.m. energy. The
  experimental data are from the INDRA collaboration.}
\end{minipage}
\end{figure}
We use the in-medium AGS equations (\ref{eqn:T3}) that reproduce
the experimental data in the limit of an isolated three-body system.
For details on the specific interaction model see
Ref.~\cite{Beyer:1996rx}. We investigate the influence of medium
dependent rates in the BUU simulation of the heavy ion collision as
compared to use of isolated (i.e. experimental) rates.
Figure~\ref{fig:PLT} shows that the net effect (gain-minus-loss) 
of deuteron production becomes larger for the
use of in-medium rates (solid) compared to using the isolated
rates (dashed). The change is significant, however, a comparison
with experimental data is difficult since deuterons may also be
evaporating from larger clusters that has not been taken into account
in the present calculation so far. The ratio of protons to deuterons
may be better suited for a comparison to experiments that is shown in
Figure \ref{fig:pd}. The use of in-medium rates (solid) lead to a
shape closer to the experimental data (dots) than the use of isolated
rates (dashed).
\begin{figure}[tb]
\epsfig{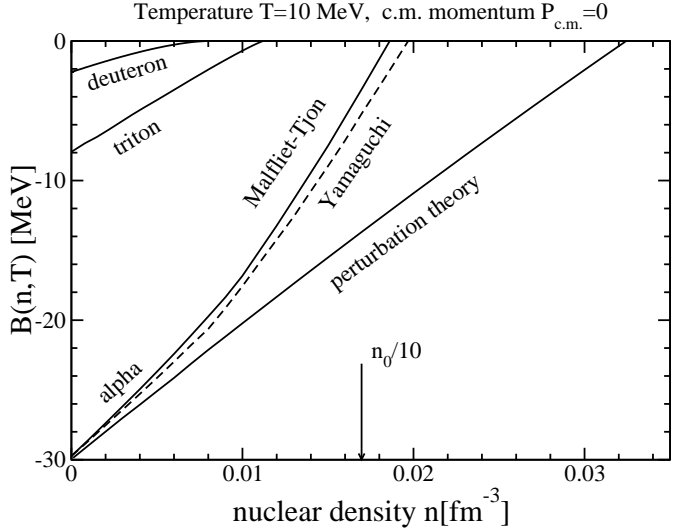}
\caption{\label{fig:amottC} 
  Difference between the pole energy of the bound state and the
  continuum, $B(n,T)=E_{\rm pole}-E_{\rm cont}$.}
\end{figure}
In these calculation, besides the change of rates, also the Mott
effect has been taken into account. 

Figure \ref{fig:amottC} shows the
dependence of the binding energy for different clusters at a given
temperature of $T=10$ MeV and at rest in the medium. 

\begin{figure}[tb]
\epsfig{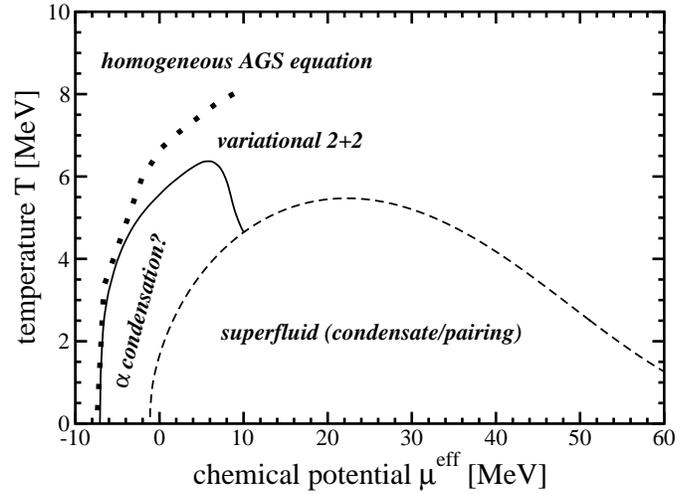}
\caption{\label{fig:Tc} 
  Critical temperatures of condensation/pairing leading to superfluid
  nuclear matter. For an explanation see text.}
\end{figure}

In Figure \ref{fig:Tc} part of the phase diagram of nuclear matter is
shown.  The condition for the onset of superfluidity for
$\ga$-particles is $B(T_c,\mu,P=0)=4\mu$. The critical temperature
found by solving the homogeneous AGS equation for $\mu<0$ confirms the
onset of $\alpha$ condensation even at higher values (dotted) as given
earlier (solid, from \cite{alpha}) which was based on a variational
calculation using the 2+2 component of the $\alpha$ particle.  For
$\mu>0$ the condition $E=4\mu$ for the phase transition can also be
fulfilled.  However, the significance for a possible quartetting needs
further investigation. Due to the many channel situation of more than
two-particles equations, the Thouless criterion might be revisited.

Work supported by Deutsche Forschungsgemeinschaft.

\end{document}